\documentclass[aps,prl,twocolumn,final,letterpaper]{revtex4}

\usepackage{graphicx}   
\usepackage{import}                         
\usepackage{amsmath} 
\usepackage{bm}
\usepackage{amssymb}
\usepackage{xspace}
\usepackage{transparent}
\usepackage{multirow}
\usepackage{mdframed}
\usepackage{color}
\usepackage{xcolor,soul}
\usepackage{bm}
\usepackage{dsfont}
\usepackage{slashed}

\usepackage{textcomp} 

\begin{document}
\rmfamily

\title{A general framework for scintillation in nanophotonics}
\author{Charles Roques-Carmes$^{1,\ddag}$}
\email{chrc@mit.edu}
\author{Nicholas Rivera$^{2,\ddag}$}
\email{nrivera@mit.edu}
\author{Ali Ghorashi$^{2}$}
\author{Steven E. Kooi$^{3}$}
\author{Yi Yang$^{1}$}
\author{Zin Lin$^{4}$}
\author{Justin Beroz$^{2}$}
\author{Aviram Massuda$^{5}$}
\author{Jamison Sloan$^{1}$}
\author{Nicolas Romeo$^{2}$}
\author{Yang Yu$^{6}$}
\author{John D. Joannopoulos$^{2,3}$}
\author{Ido Kaminer$^{7}$}
\author{Steven G. Johnson$^{2,4}$}
\author{Marin Solja\v{c}i\'{c}$^{1,2}$}

\affiliation{$^\ddag$ denotes equal contribution.\looseness=-1}
\affiliation{$^{1}$ Research Laboratory of Electronics, MIT, Cambridge, MA 02139, USA\looseness=-1}
\affiliation{$^{2}$ Department of Physics, MIT, Cambridge, MA 02139, USA\looseness=-1}
\affiliation{$^{3}$ Institute for Soldier Nanotechnologies, MIT, Cambridge, MA, 02139, USA\looseness=-1}
\affiliation{$^{4}$ Department of Mathematics, MIT, Cambridge, MA, 02139, USA\looseness=-1}
\affiliation{$^{5}$ Microsystems Technology Laboratories, MIT, Cambridge, MA 02139 USA,\looseness=-1}
\affiliation{$^{6}$ Raith America, Inc., USA,\looseness=-1}
\affiliation{$^{7}$ Department of Electrical and Computer Engineering, Technion Haifa 32000, Israel \looseness=-1}

\newpage
\begin{abstract}
    Bombardment of materials by high-energy particles (e.g., electrons, nuclei, X- and $\gamma$-ray photons) often leads to light emission, known generally as scintillation. Scintillation is ubiquitous and enjoys widespread applications in many areas such as medical imaging, X-ray non-destructive inspection, night vision, electron microscopy, and high-energy particle detectors. A large body of research focuses on finding new materials optimized for brighter, faster, and more controlled scintillation. Here, we develop a fundamentally different approach based on integrating nanophotonic structures into scintillators to enhance their emission. To start, we develop a unified and \emph{ab initio} theory of nanophotonic scintillators that accounts for the key aspects of scintillation: the energy loss by high-energy particles, as well as the light emission by non-equilibrium electrons in arbitrary nanostructured optical systems. This theoretical framework allows us, for the first time, to experimentally demonstrate nearly an order-of-magnitude enhancement of scintillation, in both electron-induced, and X-ray-induced scintillation. Our theory also allows the discovery of structures that could eventually achieve several orders-of-magnitude scintillation enhancement. The framework and results shown here should enable the development of a new class of brighter, faster, and higher-resolution scintillators with tailored and optimized performances $-$ with many potential applications where scintillators are used.
\end{abstract}

\maketitle

\section{Introduction}
Scintillation, the process by which high-energy particles (HEP, also known as ionizing radiation) bombarding a material convert their kinetic energy into light, is among the most commonly occurring phenomena in the interaction of ionizing radiation with matter. It enables a great number of technologies, including X-ray detectors used in medical imaging and non-destructive inspection, $\gamma$-ray detectors in positron-emission tomography scanners, phosphor screens in night-vision systems, electron detectors in electron microscopes, and electromagnetic calorimeters in high-energy physics experiments \cite{Gektin2017InorganicSystems, Cherry2012PhysicsMedicine}. 
Scintillation appears under many different guises. For example, when the ``high-energy" particle is a visible or UV photon, the scintillation is better known as photoluminescence. When the incident particles are energetic electrons, scintillation is also known as incoherent cathodoluminescence. 
When the high-energy particle is an X- or $\gamma$-ray, the phenomenon is almost exclusively referred to as scintillation. \cite{Gektin2017InorganicSystems}. 

Because of scintillation's broad applications, there has been (and still is) great interest in the development of ``better scintillators'' with greater photon yields, as well as greater spatial and energy resolution. Such enhanced scintillators could translate into groundbreaking functionalities. One such example is in medicine: brighter and higher-resolution scintillators could enable medical imaging (e.g., computed tomography) with higher resolution and substantially lower radiation dose, allowing early tumor screening and greater standards of care. Current approaches to improve scintillation are mostly, if not solely, oriented towards the growth of higher-quality materials (e.g., single-crystalline, controlled creation of defect sites) as well as the identification of new materials (e.g., ceramics and metal halide perovskites \cite{Chen2018All-inorganicScintillators}) with faster and brighter intrinsic scintillation.

In this manuscript, we develop a different approach to this problem, which we refer to as ``nanophotonic scintillators''. By patterning a scintillator on the scale of the wavelength of light, it is possible to strongly enhance, as well as control, the scintillation yield, spectrum, directivity, and polarization response. The contribution of our work is threefold. First, we develop a general theory of scintillation in nanophotonic structures that enables us to predict scintillation in arbitrary nanophotonic settings. Then, we use this framework to experimentally demonstrate for the first time order-of-magnitude enhancements of scintillation, as well as spectral shaping and control of scintillation. We demonstrate these effects in two different material platforms (electron scintillators and X-ray scintillators), in agreement with theoretical predictions from our framework. Third, we use the theoretical framework to discover designs that could potentially enable several orders-of-magnitude enhancement of scintillation. These contributions, taken together, should enable the systematic development of a whole new class of bright scintillators for many applications. We expect that our results will enable the systematic design of a whole new family of optimal nanophotonics-enhanced scintillators, with applications to medicine, defense, electron microscopy, and beyond.

The motivation for our approach is the observation that the light emitted in scintillation is effectively spontaneous emission \cite{Kurman2020Photonic-CrystalDetection}. An enormous amount of effort in multiple fields has gone into controlling and enhancing spontaneous emission through the density of optical states \cite{Yablonovitch1987InhibitedElectronics, Joannopoulos2011}, with corresponding impact in those fields \cite{Pelton2015ModifiedStructures}, including photovoltaics \cite{Polman2012PhotonicPhotovoltaics}, sensing \cite{Anger2006EnhancementFluorescence, Jackson2004Surface-enhancedSubstrates}, LEDs \cite{Schubert1994HighlyMicrocavities, Erchak2001EnhancedDiode}, thermal emission \cite{Greffet2002CoherentSources}, and free-electron radiation sources \cite{Remez2017, Yang2018MaximalElectrons, Kaminer2017, Roques-Carmes2019TowardsSources, Liu2017, Massuda2017, GarciadeAbajo2010, Li2016, Osorio2015Angle-ResolvedPolarimetry, Yang2021ObservationResonances}. In the context of scintillation, nanophotonic enhancements could in principle take two forms: (1) through direct enhancement of the rate of spontaneous emission by shaping the density of optical states \cite{Kurman2020Photonic-CrystalDetection}; or (2) through improved light extraction from bulk scintillators. Early pioneering work by Lecoq and coworkers demonstrated enhanced light extraction provided by a photonic crystal coating atop a bulk scintillator \cite{Liu2018EnhancedEmbossing, Knapitsch2012ResultsScintillators, Knapitsch2015ReviewScintillators, Zhu2015EnhancedLithography, Liu2012GiantNanostructures,Liu2021ImprovedCrystals, Ouyang2020EnhancedCrystals}. Nevertheless, the prospect of enhancing scintillation through the local density of states, as well as the prospect of large scintillation enhancements, by either mechanism, remains unrealized. Moreover, the type of nanophotonic structures that could even in principle realize such effects is unknown.

\begin{figure*}
    \centering
    \includegraphics[scale = 1.2]{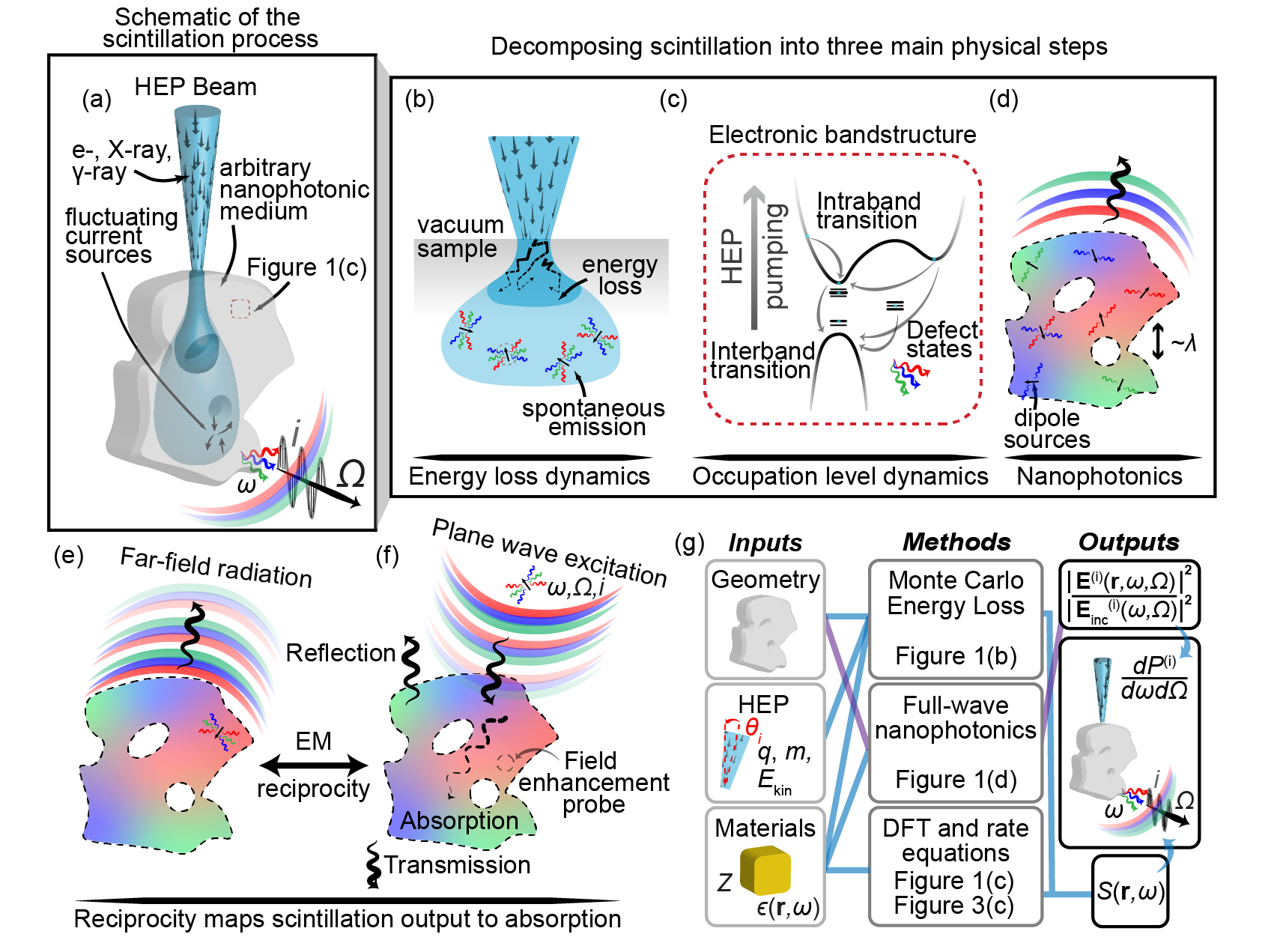}
    \caption{\textbf{A general framework for scintillation in nanophotonics.} (a) We consider the case of high-energy particles (HEP) bombarding an arbitrary nanophotonic medium, emitting scintillation photons at frequency $\omega$ (free-space wavelength $\lambda$), propagation angle $\Omega$, and polarization $i$. (b) Subsequent HEP energy loss results in excitation of radiative sites (darker blue region in sample) which may diffuse before spontaneously emitting photons (lighter blue region in sample). (c) The framework also accounts for different types of microscopic emitters. (d) The emitters may emit in arbitrary nanophotonic environments. (e-f) Electromagnetic reciprocity maps far-field radiation calculations from the stochastic many-body ensemble in a single electromagnetic simulation of plane-wave scattering, by calculating the effective spatially-dependent field enhancement. (g) Summarized framework. Links indicate forward flow of information. The purple links indicate the possibility of backward flow (inverse-design) in our current implementation. $q, m, E_\text{kin}, \theta_i$: particle charge, mass, kinetic energy, and incidence angle. $\epsilon(\mathbf{r}, \omega), Z$: material permittivity and effective $Z$-number. $S(\mathbf{r}, \omega)$: spatially-varying intrinsic scintillation spectral function. $dP^{(i)}/d\omega d\Omega$: scintillation spectral-angular power density at polarization $i$. An expanded and elaborated version of (g) is presented in the SI, Section A.}
    \label{fig:fig1}
\end{figure*}

Part of the reason for the lack of progress in this field so far entails a theoretical gap associated with the complex, multiphysics nature of scintillation emission (schematically illustrated in Figure~\ref{fig:fig1}(a-d)). The process of scintillation is composed of several complex parts spanning a wide range of length and energy scales \cite{Gektin2017InorganicSystems}: (1) ionization of electrons by HEP followed by production and diffusion of secondary electrons (Figure~\ref{fig:fig1}(b)) \cite{Klein1968BandgapSemiconductors, Polman2019}; (2) establishment of a non-equilibrium steady-state (Figure~\ref{fig:fig1}(c)) \cite{Wurfel1982TheRadiation, Greffet2018LightLaw}; and (3) recombination,  leading to light emission (Figure~\ref{fig:fig1}(d)). The final step of light emission is particularly complex to model, especially in nanophotonic settings, as it results from fluctuating, spatially-distributed dipoles with a non-equilibrium distribution function which strongly depends on the previous steps of the scintillation process. All of this together has significantly hindered direct modeling of scintillation in materials, and as of this writing, there has not been an end-to-end theoretical description of scintillation from scintillators integrated with nanophotonic structures, with potentially enhanced density of states.

\section{Results}

\subsection{A general theory of nanophotonic scintillation}

First, let us present a unified theory of nanophotonic scintillators. The theory we develop is \emph{ab initio}: it can, from first principles, predict the angle- and frequency-dependent scintillation from arbitrary scintillators (established and nascent), taking into account the three steps illustrated in Figure~\ref{fig:fig1}(b-d). It takes into account the energy loss dynamics of HEPs through arbitrary materials, the non-equilibrium steady state and electronic structure of the scintillating electrons, and the nanostructured optical environment (i.e., the electrodynamics of the light emission by this non-equilibrium electron distribution). After developing this theory, we illustrate it experimentally in two separate settings (electron-induced and X-ray induced scintillation), to show the generality of our framework. Beyond these direct applications, we also show how many of the intricate features of the scintillation (especially for electron-induced scintillation) are accounted for by the non-equilibrium kinetics and electronic structure aspects of our theory.


Consider the situation depicted in Figure~\ref{fig:fig1}(a) in which a HEP beam deposits energy into a nanophotonic structure (Figure~\ref{fig:fig1}(b)). The structure may be in proximity of a scintillating material, or integrated with it (as in both cases that we present experiments for). The interaction of the beam with the scintillating material will generally lead to a process of electron excitation in the scintillator, followed by relaxation into an excited state (Figure~\ref{fig:fig1}(c)). 

Importantly, the occupations of electrons and holes following this relaxation are typically in an approximate equilibrium \cite{Greffet2018LightLaw} (referred to as a non-equilibrium steady state). 
This equilibrium is well-defined since it occurs on picosecond timescales, which are effectively instantaneous compared to the excited state depletion timescales (nanoseconds) \cite{Klein1968BandgapSemiconductors}. 
Under these assumptions, the radiative recombination may be described in terms of emission from fluctuating currents in the material, not unlike thermal radiation (in which the electrons are in a true equilibrium). 
The key difference from thermal radiation is that the occupation functions which determine the current-current correlations (that determine the emission) are no longer governed by the Bose-Einstein distribution, but are instead material and HEP pump-dependent (and therefore spatially dependent).

Despite the non-universality of the current-current correlations, the otherwise strong similarity to thermal radiation inspires a key simplification which also gives rise to simple and powerful numerical methods for modeling and optimizing scintillation. This key simplification is electromagnetic reciprocity, which relates the following two quantities: (1) the emitted scintillation from the structure (at a given frequency $\omega$, direction $\Omega$, and polarization $i$) and (2) the \emph{absorption} of a plane wave by the scintillating structure (of frequency $\omega$, propagating along direction $\Omega$ \emph{into} the structure, and polarization $i$). As a result of this relation, it is possible to calculate the scintillation at some angle and frequency by calculating absorption of light incident from the far-field at that frequency, angle, and polarization. Direct modeling of light emission by means of calculating the emission from an ensemble of fluctuating dipoles, as considered in the past (e.g., for thermal emission \cite{Chan2006DirectSlabs}), is extremely resource-intensive from a computational perspective \footnote{This issue is compounded by the sensitivity of the results to assumptions about the spatial and spectral distributions of the dipoles, which are related to the microscopic details of the defect electronic structure, as well as mechanism of high-energy particle energy transfer into the material.}. The effect of the spatial distribution of the scintillating centers is captured by integrating this spatial distribution against the spatially-dependent absorption in the scintillating structure. In this way, the spatial information can be obtained ``all-at-once'' from a single absorption ``map''.

Let us use this simplification to quantify scintillation, which we represent in terms of the scintillation power per unit frequency $d\omega$ and solid angle $d\Omega$ along the $i$th polarization (e.g., $i=s,p$): $\frac{dP^{(i)}}{d\omega d\Omega}$ (and $\frac{dP}{d\omega d\Omega}=\sum_i \frac{dP^{(i)}}{d\omega d\Omega}$ is the total scintillation power density). In most cases, the current-current correlations in the scintillator are isotropic (a condition that we relax in the SI, Section B), and we get
\begin{equation}
    \frac{dP^{(i)}}{d\omega d\Omega} = \frac{\omega^2}{8\pi^2 \epsilon_0 c^3}\int d\mathbf{r} \frac{\left|\mathbf{E}^{(i)}(\mathbf{r},\omega, \Omega)\right|^2}{\left|\mathbf{E}^{(i)}_\text{inc}(\omega, \Omega)\right|^2} S(\mathbf{r},\omega),
    \label{eq:0}
\end{equation}
where the quantity $\mathbf{E}^{(i)}_\text{inc}(\omega, \Omega)$ denotes the electric field of an incident plane wave of frequency $\omega$, incident from a direction $\Omega$, with polarization $i$. The quantity $\mathbf{E}^{(i)}(\mathbf{r}, \omega, \Omega)$ denotes the total electric field at position $\mathbf{r}$ resulting from the incident field and their ratio is thus the field enhancement. The function $S(\mathbf{r},\omega)$ in Equation~\ref{eq:0} is the spectral function encoding the frequency and position dependence of the current-current correlations, given by $S(\mathbf{r},\omega) = \frac{1}{3}\sum_{\alpha,\beta } \text{tr}[\mathbf{J}^{\alpha\beta}(\mathbf{r})\mathbf{J}^{\beta\alpha}(\mathbf{r})]f_{\alpha}(\mathbf{r})(1-f_{\beta}(\mathbf{r}))\delta(\omega-\omega_{\alpha\beta})$. In this spectral function, $f_{\alpha}$ is the occupation factor of microscopic state $\alpha$ with energy $E_{\alpha}$, $\textbf{J}^{\alpha\beta}$ represents the matrix element of the current density operator ($\mathbf{J} \equiv \frac{e}{m}\psi^{\dagger} (-i\hbar\nabla) \psi$), $\omega_{\alpha\beta} = [E_{\alpha}-E_{\beta}]/\hbar$, and tr denotes matrix trace. Importantly, besides the position dependence of the current density matrix element, the occupation functions can also depend on position, as they depend on the HEP energy loss density (specifically, how much energy is deposited in the vicinity of $\mathbf{r}$). Interestingly, Equation~\ref{eq:0} would be proportional to the strength of thermal emission upon substitution of $S(\mathbf{r},\omega)$ by the imaginary part of the material permittivity, multiplied by the Planck function. However here, the primary difference is that $S(\mathbf{r},\omega)$ describes a non-equilibrium state, rather than the thermal equilibrium state of the material.

To better understand the core components of nanophotonic scintillation enhancement, let us simplify it further, by considering the case where the density of excited states is uniform over some scintillating volume $V_S$ (in which case we may drop the spatial dependence of $S$ such that $S(\mathbf{r}, \omega) \rightarrow S(\omega)$). This volume can be thought of as the characteristic volume over which excited electrons are created (like in Figure~\ref{fig:fig1}(b)). Then we may write
\begin{equation}
    \frac{dP^{(i)}}{d\omega d\Omega} = \frac{\pi }{\epsilon_0 \omega} \times S(\omega) \times \left[V^{(i)}_\text{eff}(\omega, \Omega)/\lambda^3\right],
    \label{eq:1}
\end{equation}
where $V^{(i)}_\text{eff}(\omega, \Omega) = \int_{V_S} d\mathbf{r}~ |\mathbf{E}^{(i)}(\mathbf{r}, \omega, \Omega)|^2 / |\mathbf{E}^{(i)}_\text{inc}(\omega, \Omega)|^2$. Having dimensions of volume, and being proportional to the absorbed power over $V_S$ (in the limit of weak absorption, so as not to perturb the field solutions), we often refer to $V^{(i)}_\text{eff}(\omega, \Omega)$ (shortened as $V_\text{eff}$) as the effective volume of field-enhancement or the effective volume of absorption. Equation~\ref{eq:1} states that the scintillation spectrum, under this approximation, is a simple product of a microscopic factor, set by the non-equilibrium steady-state distribution function $S(\omega)$, and an effective absorption volume $V_\text{eff}$, which is set only by the (structured) optical medium surrounding the scintillating medium.

Our framework to calculate scintillation according to Equation~\ref{eq:0} consists of three components, as illustrated in Figure~\ref{fig:fig1}(b-d, g): energy loss of a beam of HEPs, creation of excited electrons, and subsequent light emission (which is computed by calculating field enhancement from incident plane waves, via electromagnetic reciprocity). As a technical matter, we note that we compute the HEP energy loss density by Monte Carlo simulations of energy loss (as is standard, see Refs. \cite{Demers2011Three-dimensionalSoftware}), the electron energy levels and spectral function through density functional theory (DFT), and the nanophotonic field enhancement through finite-difference time-domain and rigorous coupled-wave analysis methods. In principle, these components are coupled together, as described in the SI, Section A.


More details on each component of the complete workflow, depicted in Figure~\ref{fig:fig1}(g) can be found in the Methods and in the SI, Sections A, B and G. The description of scintillation provided here $-$ using calculations of electronic structure, energy-loss, and electromagnetic response $-$ is to the best of our knowledge, the first to provide an \emph{ab initio} and end-to-end account of scintillation in nanophotonic structures. 

\begin{figure*}
    \centering
    \includegraphics[scale = 0.85]{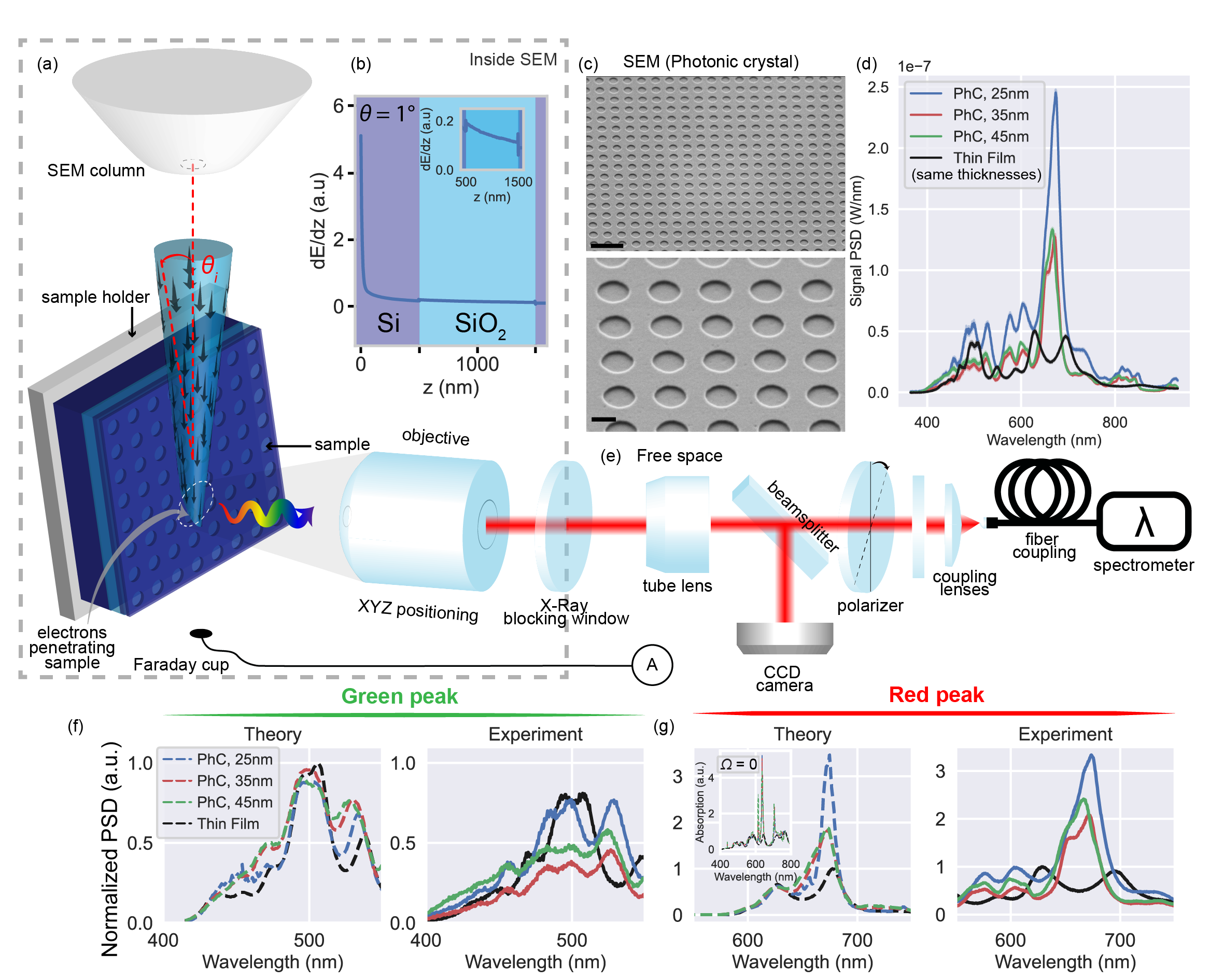}
    \caption{\textbf{Experimental demonstration of nanophotonic shaping and enhancement of electron-beam-induced scintillation .} (a) A modified scanning electron microscope (SEM) is used to induce and measure scintillation from electron beams (10-40~keV) bombarding scintillating nanophotonic structures. (b) Electron energy loss in the silicon-on-insulator wafer is calculated via Monte Carlo simulations. Inset: Zoomed-in electron energy loss in the scintillating (silica) layer. (c) SEM images of photonic crystal (PhC) sample (etch depth 35 nm). Tilt angle $45^\circ$. Scale bar: 1~$\mu$m (top), 200 nm (bottom). (d) Scintillation spectrum from thin film (TF) and PhC samples with varying etch depths (but same thickness). (e) The scintillation signal is coupled out of the vacuum chamber with an objective and then imaged on a camera and analyzed with a spectrometer. (f-g) Comparison between theoretical (left) and experimental (right) scintillation spectra for green and red scintillation peaks. Inset: Calculated scintillation spectra (per solid angle) at normal emission direction, showing the possibility of much larger enhancements over a single angle of emission.}
    \label{fig:fig2}
\end{figure*}

\section{Experimental observation of strongly enhanced scintillation}
We now present experiments demonstrating shaped and enhanced electron-beam-induced scintillation by silica defects, showing how our model accounts for features which may only be understood from a framework combining microscopic and nanophotonic details. Following this, we present experiments demonstrating enhanced X-ray-induced scintillation and imaging in cerium-doped yttrium aluminium garnet (YAG:Ce). 

\subsection{Enhanced electron-beam-induced scintillation}
We first experimentally demonstrate scintillation from silicon-on-insulator nanophotonic structures due to bombardment by electrons (here, with energies in the range of 10-40 keV). Electrons with a few tens of keV energies are a convenient platform to demonstrate nanophotonic scintillation, as they readily lose almost all of the energy to the nanophotonic structure. Such lower energy particles penetrate materials less deeply, leading to a strong overlap between the spatial region of HEP energy loss density and the region of high field-enhancement (the latter of which is within a few hundred nm of the surface). 

Our experimental setup to measure scintillation is based on a modified SEM (an earlier version of which was reported in Refs. \cite{Roques-Carmes2019TowardsSources, Yang2018MaximalElectrons, Massuda2017, Kaminer2017}), shown in Figure~\ref{fig:fig2}(a,e): a focused electron beam of tunable energy (10-40 keV) excites the sample at a shallow ($\sim 1^\circ$) angle and the resulting radiation is collected and analyzed with a set of free space optics. The light is collected by an objective lens which accepts radiation emitted in a cone of half-angle 17.5$^\circ$. Under the shallow-angle-conditions of electron incidence in our experiments, the effective penetration depth of the electrons is on the scale of a few hundred nanometers (Figure~\ref{fig:fig2}(b)), far below the nominal mean free paths of 40 keV electrons in silica or silicon, which are on the order of 20 $\mu$m. This leads to strong overlap of the energy loss with regions of field enhancement. Control over the incidence angle also enables tuning this overlap between the HEP energy loss density and $V_{\text{eff}}$. 

The first structure we consider is a thin film of 500~nm Si atop 1~$\mu$m SiO$_2$ atop a Si substrate. The second structure differs from the first in that the top Si layer is patterned to form a square lattice (design period $\sim$430 nm; see Figure~\ref{fig:fig2}(c)) of air holes (diameter $\sim$260 nm) of various etch depths ($\sim$25, 35, and 45 nm). We refer to them as ``thin film" (TF) and ``photonic crystal" (PhC) samples of same thicknesses, respectively. Scintillation in these structures occurs in the buried silica layer, and in particular, by a class of commonly occurring defects called self-trapped holes (STH) \cite{Girard2019OverviewFibers}. Such defects have been studied extensively due to their consequences for silica fibers. They display distinct emission at red and green wavelengths, which, in addition to our other observations, enable us to attribute our observations to STH defect scintillation (and thus, rule out other mechanisms of electron-beam-induced emission (such as coherent cathodoluminescence; see SI Section C)).

\begin{figure}
    \centering
    \includegraphics[scale = 0.53]{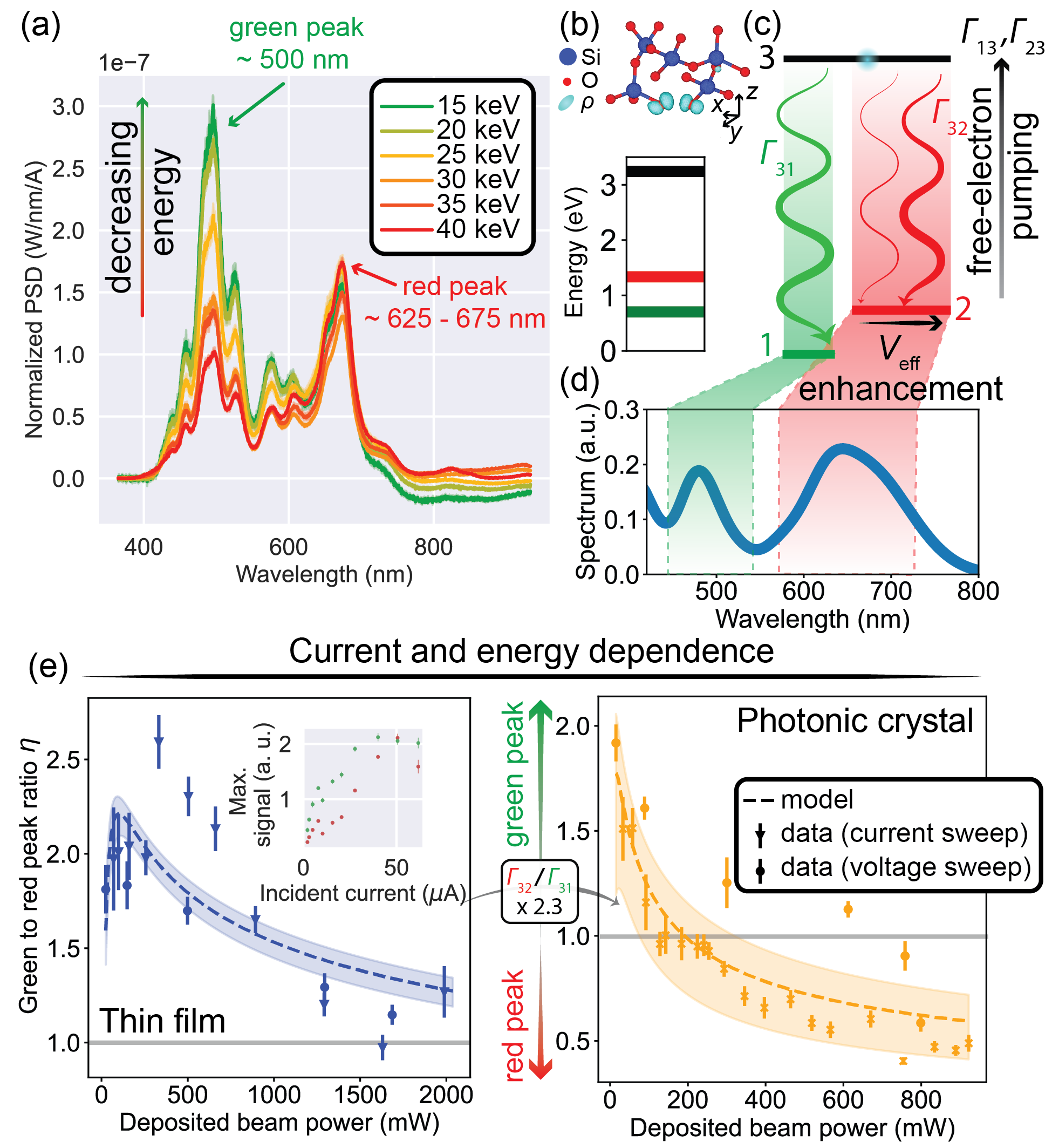}
    \caption{\textbf{Probing the microscopics of electron-beam-induced scintillation in silica.} (a) Energy-dependent scintillation spectra (PhC sample, etch 25~nm). (b) Top: 3D molecular model of STH defect in silica. Si: Silicon, O: Oxygen, $\rho$: spin-polarized density. Bottom: Calculated STH defect energy levels via density functional theory (DFT). (c) Simplified three-level system modelling the microscopics of scintillation from STH defect in silica. (d) Bulk scintillation spectrum calculated with DFT. (e) TF (left) and PhC (right) scintillation peak ratios as a function of deposited beam powers through electron pumping. The dashed line corresponds to the mean model prediction and the shaded area to the prediction from the model parameters $\pm$ their standard deviation (TF, PhC: uncertainty on  $\Gamma_{31}/\Gamma_{32}$). Inset: Maximum signal of green and red scintillation peaks versus current in TF sample.}
    \label{fig:fig3}
\end{figure}

We now show how nanophotonic structures shape and enhance scintillation in silica. The scintillation spectrum of the sample in the visible range, for both TF and PhC samples, is shown in Figure~\ref{fig:fig2}(d). The TF scintillation measurements shown in black in Figure~\ref{fig:fig2}(f, g) display two main sets of features at green ($\sim500$~nm) and red ($\sim625-675$~nm) wavelengths. At red wavelengths, there is a clear double-peak structure, while at green wavelengths, the scintillation spectrum displays multiple peaks. These multiply-peaked spectra differ considerably from prior observations of STH scintillation~\cite{Girard2019OverviewFibers}: while they occur roughly at the same wavelength, prior observations show only one peak at the red and green wavelengths \footnote{In principle, one would want to compare $V_\text{eff}$ in the TF to a “truly intrinsic” or “bulk” silica case. In that case, one would compare to silica of the same thickness (1000~nm). However, because this reference case is a thin film as well, nanophotonic shaping effects in the spectrum will inevitably be present. Comparing the $V_\text{eff}$ in the thin film case of Figure~\ref{fig:fig2} to thin films without (a) the top Si layer, and (b) without both Si layers [see SI Figure~1], one finds that the TF of Figure~\ref{fig:fig2} presents slightly smaller absorption enhancement at the red peak, possibly due to the high reflectivity of the top Si layer (suppressing the amount of field which can be absorbed by the material). However, the PhC sample still shows strong shaping and enhancement relative to all TF cases.}. The multiple peaks of the spectrum (and even its shoulders) are well accounted for at both red and green wavelengths even by the simplified Equation~\ref{eq:1}, and specifically by multiplying the shape of the STH spectrum in bulk by the $V_{\text{eff}}$ calculated for the TF. The bulk spectrum is inferred from previous observations \cite{Girard2019OverviewFibers} and confirmed by our DFT calculations (see Figure~\ref{fig:fig3}(d)). The multiply peaked structure of $V_{\text{eff}}$ thus arises from thin-film resonances, which enhance the absorption of light in the buried silica layer. The agreement between theory and experiment in Figure~\ref{fig:fig2}(f, g) unambiguously indicates a strong degree of spectral control over scintillation even in the simplest possible "nanostructure" (namely, a thin film).
%

In contrast to the TF scintillation, the scintillation from the PhC samples displays very strong and spectrally-selective enhancement. We report an enhancement of the red scintillation peak in the PhC sample, compared to the TF, by a factor of $\sim 6$ (peak at 674 nm) and of $\sim 3$ integrated over the main red peak (665 $\pm$ 30 nm) as shown in Figure~\ref{fig:fig2}(d). This feature is reproduced by our theoretical framework via enhancement of $V_\text{eff}$ around the red scintillation peak, using the same fitting parameters as those taken from the TF results of Figure~\ref{fig:fig2}(f, g). Comparatively, the green peak remains at a value similar to those in the TF spectra. Little enhancement is expected for the green wavelength, due to the high losses at those shorter wavelengths. 

The observed enhancement can readily be attributed theoretically to the presence of high-Q resonances at the red wavelength, which lead to enhanced absorption of light in the far-field. Importantly, the positions of the many subpeaks in the scintillation spectra are accounted for by the peaks of $V_\text{eff}$. Somewhat larger uncertainties are introduced in the patterned structure because of the strong degree of angular shaping of the radiation associated with certain wavevectors in the PhC bandstructure (see inset of Figure~\ref{fig:fig2}(g), showing the predicted scintillation spectrum at normal emission). As a result, the spectrum depends on the exact angular acceptance function of the objective. There is also a more sensitive dependence on the exact distribution of electron energy loss compared to the thin-film case, due to the well-localized nature of the resonances leading to scintillation in the patterned structure. 

Having shown scintillation control and enhancement based on nanophotonic structures, we move to show another core element of our general framework for scintillation:
the microscopic transition dynamics associated with the scintillation process, their effect on the non-equilibrium occupation functions, and the corresponding effect on observable properties of the scintillation spectrum. In the specific case of silica defects, we can utilize spectral observables such as dependence of the scintillation on the electron energy, as well as the ratio of green to red scintillation peak powers (defined as $\eta$) as a function of deposited HEP energy, to test assumptions about the microscopic properties of the scintillation mechanism. We can even infer the energy level structure of the scintillating defects by combining these measurements with \emph{ab initio} electronic-structure calculations and models of the excited electron kinetics (e.g., rate equations).

Figure~\ref{fig:fig3}(a) shows the evolution of the scintillation spectrum for various energies. At high-energy pumping ($\sim$40 keV), red  scintillation in the PhC sample dominates, while we observe that decreasing the pumping energy results in a gradual increase of the green peak scintillation (and of $\eta$). We took similar measurements for high and low-current pumping (at a constant pumping energy of 40 keV) of PhC and TF samples and compiled our results in Figures~\ref{fig:fig3}(e). There, one can observe that for the TF sample, the green peak scintillation always dominates ($\eta > 1$), while, for the PhC sample, there is a cross-over for a certain value of the deposited beam power (represented by $\eta$ crossing unity).

To account for these observations, we consider a description of the defect levels in terms of a three-level Fermi system, featuring two lowest occupied levels (denoted 1 and 2 in Figure~\ref{fig:fig3}(c)) coupled to an upper ``pump'' level (denoted 3) through the high-energy electron beam, which acts as a pump. These three levels correspond to energy levels from our electronic structure calculations of the STH defects in silica (based on DFT, see SI Section F). The relative rates of the transitions 3 $\to$ 1 ($\Gamma_{31}$) and 3 $\to$ 2 ($\Gamma_{32}$) -- which depend on the pump strength and the emission rates (which depend on $V_{\text{eff}}$) -- dictate the strength of the green and red emission, respectively. We arrive at the results of Figure~\ref{fig:fig3}(e) by solving for the steady-state values of these transition rates using rate equations (see Methods) and extracting the corresponding $\eta$, as a function of the incident beam power. 

The agreement between theory and experiment enables us to understand the crossover as resulting from a combination of (1) the relative enhancement of red transitions from the PhC, and (2) the nonlinear transition dynamics of excited electrons in the defect. In particular, data from both samples indicate that the pump rate for the  ``green transition'', $\Gamma_{13}$, is faster than its red counterpart, $\Gamma_{23}$ (with consistent ratio values of $\sim 3.2$ for the TF and $\sim 3.35$ for the PhC). The existence of a cross-over deposited beam power between domains where $\eta > 1$ and $\eta < 1$ translates into an enhancement of the ratio of decay rates $\Gamma_{32}/\Gamma_{31}$ in the PhC sample. Comparing model parameters fitting the TF experimental data to models fitting the PhC data, we estimate that the decay rate ratio is enhanced by a factor of $\sim 2.3 \pm 1.0$. This value is in agreement with the $V_\text{eff}$-enhancement predicted by our theory and by our observation of enhanced scintillation from the red defects in the experimental data.

By patterning nanophotonic scintillators, one can thus tailor microscopic properties and selectively enhance scintillation from microscopic defects. This also suggests that scintillation rates can be selectively enhanced using nanophotonic structures, a feature that is particularly sought after in some medical imaging modalities \cite{Lecoq2020RoadmapChallenge}. Moreover, our results indicate that the measured scintillation may be used to sort out competing models of the electronic structure, especially in complex defects such as this one, which are hard to model due to self-interaction effects (see SI Section F).

\subsection{Observation of strongly enhanced scintillation induced by X-rays}

We now move on to another example of a nanophotonic scintillator designed using our theoretical framework, showing its application to enhancing scintillation induced by high-energy photons such as X-rays. Such HEPs lose their energy much differently from massive charged particles (such as electrons). We present experimental results demonstrating enhancement of X-ray induced scintillation and corresponding X-ray image brightness enhancement generated by a common scintillator, YAG:Ce. These results immediately translate into brighter and thinner X-ray scintillators which could potentially lead to low-dose and high-resolution X-ray imaging. 

\begin{figure}
    \centering
    \includegraphics[scale = 0.7]{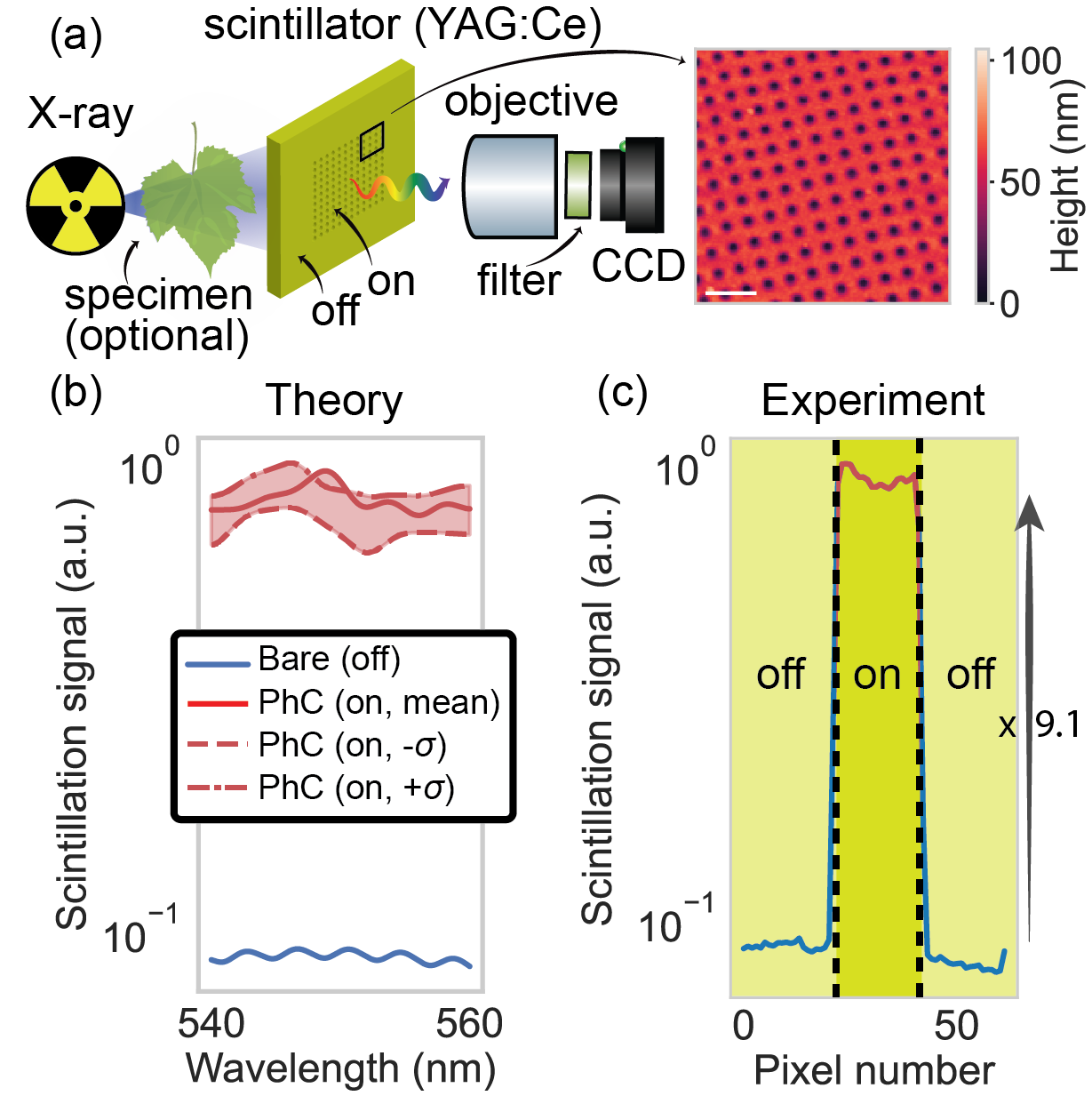}
    \caption{\textbf{Nanophotonic enhancement of X-ray scintillation.} (a, Left) X-ray scintillation experimental setup: light generated by X-ray bombardment of a cerium-doped yttrium aluminium garnet (YAG:Ce) scintillator is imaged with a set of free-space optics. A specimen may be positioned between the source and the scintillator to record an X-ray scan of the specimen. (a, Right) Atomic force microscopy image of patterned YAG:Ce scintillator (20~$\mu$m thickness). Scale bar: 1~$\mu$m. (b) Calculated scintillation spectrum of the PhC, integrated over the experimental angular aperture. Calculations are performed for measured etching depths $\pm$ a standard deviation (corresponding to 40, 50, and 60 nm). The shaded area corresponds to possible scintillation enhancements in between those values. The calculated spectra are convolved with a moving-mean filter of 1.33 nm width (raw signal shown in the SI). (c) Measured scintillation along a line of the sample, including regions on (red) and off (blue) the PhC. The scintillation from the PhC region is on average about $\times9.1$ higher than the unpatterned region. All signals were recorded with X-ray source settings: 40 kVp, 3 W.}
    \label{fig:fig4}
\end{figure}

We used the experimental configuration shown in Figure~\ref{fig:fig4}(a): X-rays traverse a specimen, leading to spatially-dependent absorption of the incident X-ray flux. This absorption pattern is geometrically magnified
until it encounters the YAG:Ce scintillator. This absorption pattern is then translated into scintillation photons which are imaged with an objective and a CCD camera. 

\begin{figure}
    \centering
    \includegraphics[scale = 0.6]{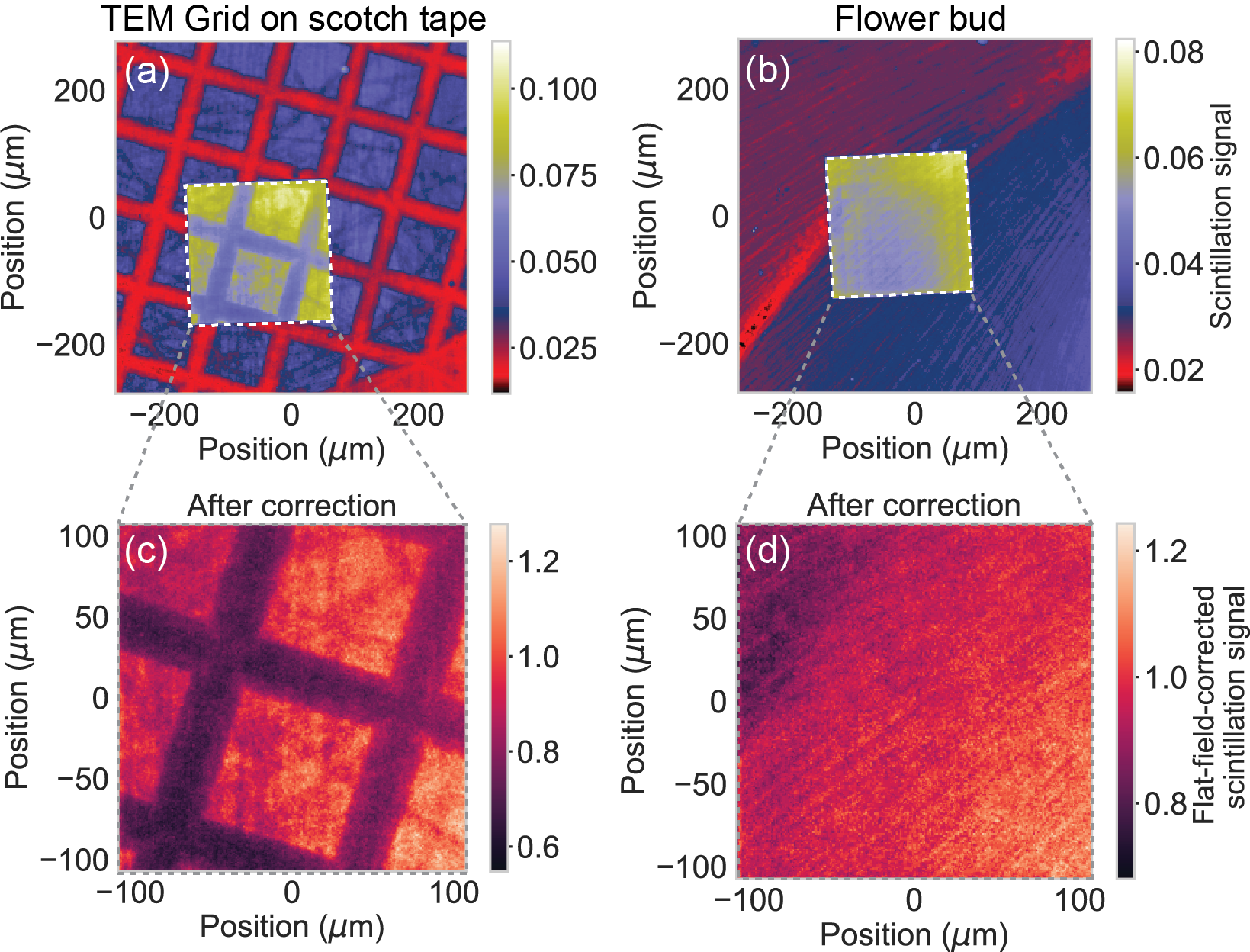}
    \caption{\textbf{X-ray scintillation imaging with nanophotonic scintillators.} (a, b) Measured X-ray images of a (a) TEM grid on scotch tape and of a (b) flower bud. The white square delimit the PhC area. (c, d) Flat-field corrected zoom-in of the X-ray image in the PhC area. Geometric magnification on those images is $\sim 2$. Compared to the unpatterned regions, the images are brighter above the PhC region, and show no evident decrease in resolution. The particular nanophotonic scintillator used for this experiment was patterned over an area of $430\times430~\mu$m and resulted in a scintillation enhancement of $\times2.3$ (measured with respect to unpatterned scintillator of same thickness). All signals were recorded with X-ray source settings: 60 kVp, 5 W.}
    \label{fig:fig5}
\end{figure}

The nanopatterned scintillator is constructed by etching a two-dimensional PhC into YAG (via Focused Ion Beam (FIB) lithography; see Methods), at the surface of the scintillator facing the objective. The PhC period is 430~nm and the total patterned area is 215~$\mu$m~$\times~$215~$\mu$m (in Figure~\ref{fig:fig4}) or 430~$\mu$m~$\times~$430~$\mu$m (in Figure~\ref{fig:fig5}). Let us now apply our general framework to predict the enhancement of scintillation that a PhC could provide.

In the case of YAG:Ce, the intrinsic scintillation properties have been long characterized and our experiments reveal only weak dependence of the scintillation on incident X-ray energy (see influence of X-ray filter in SI, Section H). Thus, the full theoretical apparatus we demonstrate for electron scintillation is not needed to adequately describe our results. Primarily, the electromagnetic response (using reciprocity) is needed to account for the experimental results, and is the  part of our general framework that leads us to order-of-magnitude enhancement of X-ray scintillation. We discuss in greater detail our numerical methods and comparison to experiment in the SI, Section H. 

According to the scintillation framework developed in the previous sections, nanophotonic scintillation enhancement is to be expected when the absorption of light is enhanced. In Figure~\ref{fig:fig4}(b) we show the calculated wavelength-dependent scintillation in YAG:Ce (averaged over the angular acceptance of the objective, as in Figure~\ref{fig:fig2}) for an unpatterned self-standing thick (20~$\mu$m) film, as well as for the PhC sample. 
Here, the calculated enhancement is by a factor of $\sim 9.3\pm 0.8$ over the measured scintillation spectrum. In our calculations, we attribute the main error bar to the uncertainty on the hole depth ($\pm 10$~nm as can be extracted from our AFM measurements, shown fully in Figure~\ref{fig:fig4}(a, Right) and in cross-sections in the SI). However, we should note that there are several other sources of uncertainty in the fabricated samples: the hole diameter and periodicity, and the optical absorption of YAG:Ce (taken in our calculations to be the value provided by the wafer supplier). We also measured and compared to our theory scintillation enhancements from multiple nanophotonic scintillators with various thicknesses, hole shapes, depths and patterned areas (see additional experimental data and Table~I in the SI).

Here, the X-ray scintillation enhancement originates in light out-coupling enhancement (or by reciprocity, in-coupling enhancement). In particular, the PhC allows more channels (i.e. a plane-wave coupling to a resonance) into the scintillator crystal, compared to a flat interface. The multiple channels translate into sharp resonant peaks in the calculated absorption spectrum (raw signal shown in the SI, Section H). This is to be contrasted with the origin of electron-beam-induced scintillation enhancement in silica, where the enhancement can be tied to the presence of a single, or small number of high-Q resonances. This effect is of the type often leveraged to design more efficient LEDs and solar cells that approach the ``Yablonovitch limit'' in both ray-optical \cite{Yablonovitch1982StatisticalOptics, Campbell1986TheSunlight}, and nanophotonic \cite{Raman2010FundamentalStructures, Yu2010FundamentalCells} settings. There, it is well known that the device efficiency is optimized by designing a structure that leads to strong absorption over the spectral range of the emission \cite{Yablonovitch1982StatisticalOptics, Zhmakin2011EnhancementDiodes}. 

In Figure~\ref{fig:fig4}(c), we show the experimentally measured scintillation scanned along a line of the sample. The regions ``off'' indicate unpatterned regions of the YAG:Ce, while ``on'' indicates the PhC region. Here, the signal is enhanced on average by a factor of $\sim 9.1$ over the unpatterned region, consistent with the predictions of Figure~\ref{fig:fig4}(b).  

To demonstrate the potential of our approach to X-ray imaging, we fabricated a larger-scale pattern on a 50~$\mu$m wafer which exhibits a scintillation enhancement of 2.3 (see measurement and theory in SI, Section H). We recorded single-shot X-ray scans of biological and inorganic specimens through the PhC, showing no evident decrease in resolution, while increasing the image brightness by the same factor. Equivalently, the required X-ray dose or exposure time to get a given number of counts on the detector is reduced, as is shown experimentally in the SI, Section H. 


Our framework allows us to further gain understanding of the scintillation mechanism at play, directly leveraging known techniques in absorption enhancement. One could expect even greater scintillation enhancements on the order of $\sim 4n^2$ in the ray-optics approximation \cite{Yablonovitch1982StatisticalOptics} or $\sim 4 \pi n^2$ for periodic structures on the wavelength scale \cite{Raman2010FundamentalStructures, Yu2010FundamentalCells} (where $n$ is the index of refraction). For example, for a high-index material such as doped GaAs, which also scintillates at room temperature \cite{Derenzo2021HowGaAs}, enhancements on the order of $\sim 50$ and $\sim 150$ could be respectively achieved in the two regimes (over a $2\pi$ collection solid angle).


\section{Discussion}

We have presented a general framework to model, tailor, and enhance scintillation by means of nanophotonic structures integrated into scintillating materials (nanophotonic scintillators). While we mainly focused on the demonstration of spectral shaping and enhancement of scintillation, our results could be extended to show angular and polarization control as well. We have demonstrated nanophotonic scintillators enhancing electron-beam-induced and X-ray-induced scintillation. The theoretical framework we used to describe our experimental results combines Monte Carlo simulations of the energy loss density \cite{Demers2011Three-dimensionalSoftware} with DFT calculations of the microscopic structure and full-wave calculations of the electromagnetic response of the nanophotonic structures probed in this work. 

We note that this type of ``full'' analysis has to the best of our knowledge not been performed to explain scintillation (nor incoherent cathodoluminescence) experiments, likely due to the prohibitively expensive computations associated with simulating ensembles of dipoles radiating in 3D structures. The reciprocity framework we use (also commonly used in areas of thermal radiation, LEDs, and photoluminescence \cite{Greffet2018LightLaw, Liu2018Light-EmittingRadiation, Schulz2018ReciprocitySystem, Sheng2011, Zhang2015CalculationTheorem, Overvig2021ThermalInteractions}) strongly simplifies the analysis, and makes a full modeling of the scintillation problem tractable. We conclude by outlining a few promising avenues of future work that are enabled by the results provided here. Further elaboration and initial results, for each of these avenues, is detailed in the SI.

The first area, inspired by our simplified calculations based on reciprocity, is numerical optimization of nanophotonic scintillators. Our framework, which relies on the calculation of $V_\text{eff}$ (which is relatively amenable, even in 3D), enables the inverse-design of nanophotonic scintillators. Methods to calculate the forward ($V_\text{eff}$ given a nanophotonic structure) and backward (gradients of $V_\text{eff}$ with respect to degrees of freedom describing the nanophotonic structure) problems are detailed in the SI, Section E. The experimentally reported enhancements can be further improved upon by inverse-designing the nanophotonic structure via topology optimization of $V_\text{eff}$ \cite{Molesky2018InverseNanophotonics}. In the SI, we show the kind of results that could be expected from topology-optimized nanophotonic scintillators: we find that 10-fold, and even nearly 100-fold selective enhancements of scintillation in topology-optimized photonic structures are possible. By considering different emission linewidths and frequencies, one can selectively design optimized nanophotonic structures which enhance one of the scintillating peaks, at a single-frequency or over the entire scintillation bandwidth. Beyond our reciprocity-based approach, low-rank methods can be utilized for the inverse-design of nanophotonic scintillators with very large angular ranges \cite{Polimeridis2015FluctuatingGeometries}. Beyond scintillation, our techniques may find applicability in other imaging modalities involving random incoherent emitters, such as surface-enhanced Raman scattering \cite{Christiansen2020InverseScattering}.

Another promising area of research enabled by our findings is nanophotonically-enhanced and -controlled UV light sources. In the SI, Section F, we show how UV scintillation in materials such as hBN enables strongly enhanced scintillation with a spectrum that can be controlled simply by the position of the electron beam relative to the patterned features in the hBN arising from changes in the overlap between the HEP loss density and $V_{\text{eff}}$. 
The prospect of realizing optimized and compact nanophotonic UV scintillation sources is particularly exciting for applications in water purification and sanitization \cite{Watanabe2009}.

To summarize, nanophotonic scintillators provide a versatile approach for controlling and enhancing the performance of scintillating materials for a wide range of applications. The framework developed here applies to arbitrary scintillating materials, nanophotonic structures, and HEPs, solving for the process end-to-end using first-principles methods. The electron-beam and X-ray scintillation experiments provide the proof-of-concept tests of the promising prospects of this field. Our work may open a panoply of exciting applications, from high-resolution, low-dose X-ray imaging to efficient ultraviolet electron-beam-pumped light sources. 


\newpage

\section{Summary of methods and materials}
\subsection{Experimental} \textit{Electron-beam experiments.} We use a modified CamScan CS3200 custom Scanning Electron Microscope (SEM) from Applied Beams (Oregon). The electron emitter is a LaB6 emitter cathode operated with settings producing the highest currents (typically $>$ 20 $\mu$A). Measurements are performed at the highest magnification (equivalent to spot mode). The sample is mounted on a 6-axis, fully eucentric stage, at a working distance of about 70 mm. 

A Nikon TU Plan Fluor ×10 objective with a numerical aperture (NA) of 0.30 was used to collect light from the area of interest. The spectrometer used was an Acton SP-2360–2300i with a low-noise, deep-cooled PIXIS camera. Monochrome images of the radiation were collected with a Hamamatsu CCD, in order to align the optical setup and spatially resolve the observed radiation. 

The objective is mounted on a 5-axis (XYZ, two tilt angles) homemade positioning stage. The focal spot of the objective is aligned with the electron beam focus (and sample surface). Two piezoelectric motors allow the objective to move in a plane parallel to the sample surface. A compact motorized actuator controls the distance of the objective to the sample surface. Two additional manual adjustment knobs allow control the alignment of the objective focal plane with the sample surface. The current is measured through a Faraday cup in the SEM stage, connected to a Keathley 6485 picoammeter. The picoammeter is triggered to acquire current signals during a time interval corresponding to the optical acquistion time (10 averaged acquisitions of 1 second duration, unless otherwise specified). A calibration measurement is performed with a calibrated light source of known power spectral density to convert the measured spectra to absolute power spectral densities and efficiencies. More information on the experimental setup can be found in the Supplementary Information, Section B. All spectra recorded with the spectrometers were averaged over 10 acquisitions of 1 second each. 

\textit{X-ray experiments.} Experiments were carried inside the enclosure of a ZEISS Xradia 520 Versa micro-CT machine. The same objective (Nikon TU Plan Fluor ×10) was mounted on the detector stage, and positioned to record an image of the surface of the scintillator. The scintillator and specimen were mounted on the same sample stage. Visible filters were taped directly at the back of the objective. In the images shown in Figure~\ref{fig:fig4} and \ref{fig:fig5}, no X-ray filters ("Air" setting) and a narrow bandpass visible filter (AVR Optics FF01-549-15-25) were used. Additional data showing the influence of visible and X-ray filters is given in the SI, Section H.

\subsection{Sample} The sample wafer for electron-beam-induced scintillation was purchased from MEMS Material and Engineering, Inc. (Sunnyvale, CA). The wafer was fabricated by a fusion bonding - grinding - polishing process. The wafer is made of a device layer (\textit{p}-doped polished silicon, $\langle100\rangle$ orientation, resistivity 1-30 $\Omega$.cm, thickness 0.5 $\pm$ 0.025 $\mu$m), on top of an oxide layer (amorphous silica, thickness 1.0 $\mu$m $\pm$ 5$\%$), on top of a handle wafer (\textit{p}-doped silicon, $\langle100\rangle$ orientation, resistivity 1-30 $\Omega$.cm, thickness 625 $\pm$ 10 $\mu$m). The patterning was produced by Dr. Timothy Savas with optical interference lithography. The YAG:Ce crystal used in the X-ray experiment was purchased from Crytur and patterned with a VELION FIB-SEM. Fabrication parameters are given in the SI, Section H. One reason we employed the VELION FIB-SEM is that nanofabrication techniques to pattern YAG:Ce are limited. Another reason is that the VELION’s FIB field has astigmatism and distortion corrections, enabling more accurate large-area FIB patterning. Finally, we selected the Au$^{+}$ FIB because it conveniently matched the Au later that would subsequently be removed with Au selective etchant.

\subsection{Fitting to experiments} 
\textit{Electron-beam-induced scintillation.} The experimentally obtained spectra in Figure~\ref{fig:fig2}(d) were accounted for based on Equation~\ref{eq:0} of the main text. The red and green peaks of STH were separately fitted (hence, no assumption is made about the relative oscillator strengths of the two peaks). The spectral dependence of $S(\mathbf{r}, \omega)$ was taken as a sum of two Gaussians at the red and green peaks, on account of inhomogeneous broadening of the defect levels. Fits were obtained taking the red and green peak energies to be 1.95 and 2.6 eV respectively, with respective FWHM of 0.25 eV and 1.2 eV. Both the peak energies and widths are consistent with previous experimental measurements of STH spectra \cite{Girard2019OverviewFibers}, as well as with our DFT calculations. The function $V_{\text{eff}}$, as defined in Figure~\ref{fig:fig1} is calculated using rigorous coupled-wave analysis.

The function $V^{(i)}_{\text{eff}}(\omega,\Omega)$ is calculated through the volume-integrated field enhancement of a plane wave incident from the far-field at angles $\Omega=(\theta,\phi)$ with polarization $i \in \{s,p\}$ and frequency $\omega$. The integration volume (particularly, the effective depth inside silica) is fitted to provide a good agreement with experiment, and accordingly the integrand of $V_{\text{eff}}$ is integrated to a depth of 500 nm inside the silica layer, which is within a factor of 2 of the effective depth predicted from CASINO and is within the uncertainty of the incident angle of the electron beam. The theoretically predicted signals are averaged over the numerical aperture of the objective (17.5$^{\circ}$) and summed over polarizations.  The data is best explained assuming that the samples have a small ($\sim 8^{\circ}$) misalignment of their normal to the axis of the objective, with the 25 nm sample oppositely oriented from the other samples.

\textit{X-ray-induced scintillation.} Absorption maps are calculated with rigorous coupled-wave analysis, with geometrical parameters extracted from SEM/AFM measurements. The reported value of the loss in the unpatterned YAG:Ce film is of $\text{Im}(\epsilon) \sim \times10^{-6}$ (information provided by Crytur). Geometrical parameters are extracted via an atomic force microscopy measurement fitted to as $\sin^2$ profile. Error bars on the predicted enhancements are calculated by varying the geometrical parameters according to the measured error bars from the characterization.   

\subsection{Monte Carlo HEP Energy Loss Simulations} 
HEP energy loss was calculated for energetic free-electrons impinging on the (unpatterned) silicon-on-insulator wafer using the open source CASINO Monte Carlo software \cite{Demers2011Three-dimensionalSoftware}. Calculations of the position-dependent energy loss density, $\frac{dE}{dV}(x,y,z)$ were done for electrons incident at shallow angles of incidence ($\sim1^\circ$ measured with respect to the substrate plane) by averaging over results from 250,000 incident electrons. The data was used to calculate the marginal electron energy loss distribution per depth $\frac{dE}{dz} = \int dx dy ~ \frac{dE}{dV}$ shown in Figure~\ref{fig:fig2}(b). We note that these calculations were also used to model scintillation in patterned samples, thus effectively neglecting the influence of the shallow pattern on the electron energy loss map. 

Calculations were also performed to find the energy loss density as a function of the incident electron energy, which was used as input in the fits of Figure \ref{fig:fig3}(e).  Similar calculations were also done for predictions of enhanced luminescence of boron nitride in the SI, Section F.

\subsection{Density Functional Theory (DFT) Calculations} 

DFT calculations \cite{Sundararaman2017JDFTx:Theory, Freysoldt2009DirectTheory} were performed on one bulk and three cluster models of STH. Cluster calculations used the Boese-Martin exchange correlation functional with 42\% exact exchange \cite{Boese2004DevelopmentKinetics} to take into account self-interaction effects. Dangling bonds were passivated with hydrogen atoms to mitigate their effect on the electronic structure. A 20 Hartree plane wave cutoff was used and Coulomb truncation \cite{Sundararaman2013RegularizationSystems} was implemented to mitigate the effects of cluster-cluster interactions. The defect transitions observed were attributed to localized states at the oxygen atoms -- verified by calculations of the spin density. 

Bulk models, shown in Figure~\ref{fig:fig3}, with constrained 1 Bohr/unit cell magnetization yielded trapped hole defects without the need for hybrid functionals. These models yielded the same transition energies as above but used the PBE exchange correlation functional \cite{Perdew1996GeneralizedSimple}. Additional details on the various DFT models and calculation results are shown in the SI, Section G.

\subsection{Three-level rate equation model} 
Based on DFT calculations, a simplified three-level system is designed to model electron pumping and subsequent radiative emission from defect states in silica. The model is pictured in Figure~\ref{fig:fig3}(c), corresponding to calculated energy levels from the DFT model in Figure~\ref{fig:fig3}(a, bottom). The following rate equations are used to model the system:
\begin{equation}
    \begin{cases}
      \frac{dp_1}{dt} =&  - \Gamma_{13}~p_1 (1-p_3) + \Gamma_{31}~p_3 (1-p_1)\\
      \frac{dp_2}{dt} =&  - \Gamma_{23}~p_2 (1-p_3) + \Gamma_{32}~p_3 (1-p_2)\\
      \frac{dp_3}{dt} =&  \Gamma_{13}~p_1 (1-p_3) - \Gamma_{31}~p_3(1-p_1)\\  
      &+ \Gamma_{23}~p_2 (1-p_3) - \Gamma_{32}~p_3 (1-p_2)
    \end{cases}
    \label{eq:rate-eq-model}
  \end{equation}
such that the total occupation probability is conserved over time $\frac{d\sum_i p_i}{dt} = 0$ with the initial condition $p_1 = p_2 = 1$ and $p_{3} = 0$. This set of equations describe a three-level system, where $1$ (resp. $2$) is the ground state corresponding to green (resp. red) emission, $3$ is a shared excited state to which electrons are sent via free-electron pumping. Band electrons can relax from the excited state $3$ to one of two ground states $1$ and $2$, corresponding to the green and red peak emission, respectively. 

We can solve the steady-state of equation~\ref{eq:rate-eq-model} to estimate the ratio of green to red emission at the steady-state:
\begin{equation}
    \eta = \frac{\Gamma_{31}~(1-p_1)}{\Gamma_{32}~(1-p_2)}.
\end{equation}
Calculations were performed using the DifferentialEquations.jl package in Julia \cite{Rackauckas2017DifferentialEquations.jlJulia} and fit using the LsqFit.jl package.

We use this model to gain further microscopic understanding of the observed experimental data, in conjunction with the general nanophotonic scintillator theory described in the main text. We chose $\eta$ as an experimental observable, since it can be calculated from equation~\ref{eq:rate-eq-model} and -- assuming green and red peak defects are localized in the same region -- the observable is independent of a few experimental unknowns (beam size, number of excited emitters). Electrons in state 3 can then radiatively decay into state 1 or 2.

We assume that $\Gamma$ is proportional to the electron beam energy deposited in the luminescent material: $\Gamma \propto I\times E \times \eta_\text{ene}(E)$ where $I$ is the incident electron current, $E$ its acceleration voltage, and $\eta_\text{ene}(E)$ the fraction of energy (normalized to the incident energy $E$) deposited by an electron in the silica layer, calculated via Monte-Carlo Simulations of electron scattering in the TF sample \cite{Demers2011Three-dimensionalSoftware} (see corresponding Methods section).

In a first numerical experiment shown in Figure~\ref{fig:fig3}(e, left), we utilized scintillation data measured on the TF sample at various incident voltages and currents. This data was used to estimate the ratio of pumping rates $\Gamma_{13}/\Gamma_{23} = 3.2 \pm 0.09$. This values indicates an intrinsic preference of the system to excite the green defect through electron pumping.

In a second numerical experiment shown in Figure~\ref{fig:fig3}(e, right), we utilized scintillation data measured on the PhC sample at various incident voltages and currents. This data was used to estimate the ratio of decay rates enhancements $\Gamma_{32}/\Gamma_{31}$ and to confirm the value of $\Gamma_{13}/\Gamma_{23}$. When letting both parameters be optimized, we obtain a value of $\Gamma_{13}/\Gamma_{23} = 3.35 \pm 0.13$, similar to the original value. We can also estimate the value of $\left(\frac{\Gamma_{32}}{\Gamma_{31}}\right)_\text{PhC}\left(\frac{\Gamma_{32}}{\Gamma_{31}}\right)_\text{TF}^{-1} \sim 2.3$ which corresponds to the scintillation rate enhancement of the red defects. This value is in agreement with our calculations and experimental demonstration of $V_\text{eff}$ scintillation enhancement of the red defects. The relative error on this estimate is of $\pm 0.4$ (uncertainty coming from the first numerical experiment) and of $\pm 0.9$ (uncertainty coming from the second numerical experiment). Therefore, results from the three-level model are a strong indication of the microscopic nature of the observed scintillation spectrum. 

We verified the robustness of our fits by trying different differential equation solvers and fitting methods, and did not observe any significant change in the values obtained for the parameters of interest, which indicates the consistency of our approach. For instance, another local optima of the optimization, which we did not detail for the sake of brevity, had the following parameters: $\Gamma_{13}/\Gamma_{23} = 4.43 \pm 0.94$ (TF data only), $\Gamma_{13}/\Gamma_{23} = 4.42 \pm 0.17$ (PhC data only), and $\left(\frac{\Gamma_{32}}{\Gamma_{31}}\right)_\text{PhC}\left(\frac{\Gamma_{32}}{\Gamma_{31}}\right)_\text{TF}^{-1} \sim 4.06$, with relative error on this estimate of $\pm 1.42$ (uncertainty coming from the second numerical experiment) and of $\pm 7.15$ (uncertainty coming from the first numerical experiment). Though the error bar in Figure~\ref{fig:fig3} only shows the relative model uncertainty with respect to the value of $\frac{\Gamma_{32}}{\Gamma_{31}}$ (which is the main decay rate variable relating to our experimental observables), we observe that the relative error on other parameters is comparable or lower.

\section{Authors contributions}
C.~R.-C., N.~Ri., N.~Ro., I.~K., and M.~S. conceived the original idea. N.~Ri. developed the theory with inputs from C.~R.-C. and A.~G. C.~R.-C. and S.~E.~K. performed the electron-beam and X-ray experiments. C.~R.-C. and N.~Ri. analyzed the experimental data and fitted it to the theory. C.~R.-C. and S.~E.~K. built the electron-beam experimental setup with contributions from J.~B., A.~M., J.~S., Y.~Ya., I.~K., and M.~S. ~ N.~Ri. performed energy loss calculations. C.~R.-C. performed absorption map calculations. A.~G. performed DFT calculations. C.~R.-C. wrote code for optimizing nanophotonic scintillators with inputs from N.~Ri., Z.~L. and S.~G.~J. Y.~Yu and C.~R.-C. fabricated the X-ray scintillation sample. ~ J.~D.~J., I.~K., S.~G.~J., and M.~S. supervised the project. The manuscript was written by C.~R.-C. and N.~Ri. with inputs from all authors. 

\section{Competing interests}
The authors declare the following potential competing financial interests: C.~R.-C., N.~Ri., A.~G., S.~E.~K., Y.~Y., Z.~L., J.~B., N.~Ro., J.~D.~J., I.~K., S.~G.~J., and M.~S. are seeking patent protection for ideas in this work (Provisional Patent Application No. 63/178,176). C.~R.-C., N.~Ri., Z.~L., and M.~S. are seeking patent protection for ideas in this work (Provisional Patent Application No. 63/257,611).

\section{Data and code availability statement}
The data and codes that support the plots within this paper and other findings of this study are available from the corresponding authors upon reasonable request. Correspondence and requests for materials should be addressed to C.~R.-C. (chrc@mit.edu) and N.~Ri. (nrivera@mit.edu).

\section{Acknowledgements}
The authors thank Tim Savas for assistance in fabricating the sample used for electron-beam scintillation. The authors thank Irina Shestakova and Olivier Philip (Crytur) for helpful discussions on X-ray scintillators; Christoph Graf vom Hagen, Xiaochao Xu and John Treadgold (Zeiss) for feedback on micro-CT scanner experiments; Ravishankar Sundararaman (Rensselaer Polytechnic Institute) 
and Jenny Coulter (Harvard University) for assistance with DFT calculations; Yannick Salamin, and Simo Pajovic (MIT) for stimulating discussions.
This material is based upon work supported in part by the U.S. Army Research Laboratory and the U.S. Army Research Office through the Institute for Soldier Nanotechnologies, under contract number~W911NF-18–2–0048. 
This material is also in part based upon work supported by the Air Force Office of Scientific Research under the award number FA9550-20-1-0115, as well as in part supported by the Air Force Office of Scientific Research under the award number FA9550-21-1-0299.
This work was performed in part on the Raith VELION FIB-SEM in the MIT.nano Characterization Facilities (Award: DMR-2117609)
C.~R.-C. acknowledges funding from the MathWorks Engineering Fellowship Fund by MathWorks Inc. 
\bibliographystyle{ieeetr}
\bibliography{references.bib}
\end{document}